\documentclass[conference]{IEEEtran}
\IEEEoverridecommandlockouts
\usepackage{cite}
\usepackage{amsmath,amssymb,amsfonts}
\usepackage{algorithmic}
\usepackage{graphicx}
\usepackage{textcomp}
\usepackage{xcolor}
\usepackage{comment}
\usepackage{cuted}
\usepackage{url}
\usepackage[nolist,nohyperlinks]{acronym}

\begin{acronym}[AAAAAAAAAAAAAAAAAAAAAAAA]  % longest acronym to fix width
 \acro{3GPP}{third generation partnership project}
 \acro{4G}{fourth-generation}
 \acro{5G}{fifth-generation}
 \acro{6G}{sixth-generation}
 \acro{BS}{base station}
 \acro{CDF}{cumulative distribution function}
 \acro{CoMP}{coordinated multi-point}
 \acro{CSI-RS}{channel state Information reference signal}
 \acro{dB}{decibel}
 \acro{dBi}{decibel isotropic}
 \acro{DRL}{deep reinforcement learning}
 \acro{DFT}{discrete Fourier transform}
 \acro{FR}{Frequency Range}
 \acro{FSO}{free-space optical}
 \acro{HAPS}{high-altitude platform station}
 \acro{HPBW}{half-power beamwidth}
 \acro{IoT}{internet of things}
 \acro{IRS}{intelligent reflecting surfaces}
 \acro{ISD}{inter-site distance}
 \acro{LAPS}{low-altitude platform station}
 \acro{LoS}{line-of-sight}
 \acro{LTE}{long term evolution}
 \acro{MIMO}{multiple-input multiple-output}
 \acro{mMIMO}{massive Multiple Input/Multiple Output}
 \acro{mmWave}{millimetre wave}
 \acro{NLoS}{non-line-of-sight}
 \acro{NR}{new radio}
 \acro{NTN}{non-terrestrial network}
 \acro{O2I}{outdoor-to-indoor}
 \acro{PRB}{physical resource block}
 \acro{RF}{radio frequency}
 \acro{RL}{reinforcement learning}
 \acro{RSRP}{reference signal received power}
 \acro{SINR}{signal-to-interference-plus-noise ratio}
 \acro{SNR}{signal to noise ratio}
 \acro{SSB}{synchronization signal block}
 \acro{TBS}{terrestrial base station}
 \acro{THz}{terahertz}
 \acro{TN}{terrestrial network}
 \acro{TUAV}{tethered unmanned aerial vehicle}
 \acro{UAV}{unmanned aerial vehicle}
 \acro{UE}{user equipment}
 \acro{UMa}{urban macro}
 \acro{UPA}{uniform planar array}
 \acro{QoS}{quality of service}
\end{acronym}

\usepackage[normalem]{ulem}

\usepackage{pgfplots}
\usetikzlibrary{patterns}
\pgfplotsset{compat=newest}

\usepackage{booktabs}
\usepackage{tabularx}

\def\BibTeX{{\rm B\kern-.05em{\sc i\kern-.025em b}\kern-.08em
    T\kern-.1667em\lower.7ex\hbox{E}\kern-.125emX}}

\DeclareMathOperator*{\argmax}{\arg\!\max}

\renewcommand{\figurename}{Figure}

\begin{document}

\title{Low-Altitude Platform Stations for Dense Urban Connectivity}
\author{
\IEEEauthorblockN{
German Svistunov\IEEEauthorrefmark{1},
Azim Akhtarshenas\IEEEauthorrefmark{1}\IEEEauthorrefmark{2},
Matteo Bernabè\IEEEauthorrefmark{1},
and David López-Pérez\IEEEauthorrefmark{1}%\IEEEauthorrefmark{2}
}
\normalsize\IEEEauthorblockA{\emph{\IEEEauthorrefmark{1}iTEAM Research Institute, Universitat Politècnica de València (UPV), Spain}\\
\IEEEauthorrefmark{2}Eindhoven University of Technology (TU/e), Eindhoven, The Netherlands}
% \normalsize\IEEEauthorblockA{\emph{\IEEEauthorrefmark{2}Beihang Valencia Polytechnic Institute (BVPI), China}}
% 

\thanks{
This research is supported by the Generalitat Valenciana, Spain, through the CIDEGENT PlaGenT, Grant CIDEXG/2022/17, Project iTENTE, and the action CNS2023-144333, financed by MCIN/AEI/10.13039/501100011033 and the European Union “NextGenerationEU”/PRTR.} 
}
\maketitle

%%%% Abstract
\begin{abstract}
This paper evaluates the downlink performance of 5G non-terrestrial networks (NTNs) realized via low-altitude platform station (LAPS)-mounted base stations, and compares it against conventional 5G terrestrial network (TN) in a realistic dense urban scenario. 
% Unlike battery-limited UAVs, TUAVs are connected to ground stations 
% via lightweight cables, enabling stable positioning and near-line-of-sight links to 
% users without the endurance constraints of untethered platforms. 
Using a 3GPP-compliant multi-cell hexagonal layout with 19 sites and 57 sectors, we model LAPS altitudes ranging from 100 to 1000\,m and evaluate the average effective signal-to-interference-plus-noise ratio and user throughput via system-level simulations. Results show that LAPS-based NTN deployments can significantly outperform terrestrial 5G in per-user throughput, provided the LAPS altitude is properly chosen to balance increased line-of-sight probability against inter-beam interference.
\end{abstract}

% \begin{IEEEkeywords}
% non-terrestrial network, HAPS, network performance, SINR, 3GPP 
% \end{IEEEkeywords}
%\noindent\textit{This work has been submitted to the IEEE for possible publication. Copyright may be transferred without notice, after which this version may no longer be accessible.}
%%% Introduction
\section{Introduction}

The rapid growth of mobile data traffic and emerging applications has imposed unprecedented demands on \acp{TN} in terms of coverage and capacity~\cite{azari2022evolution}. 
To address these demands, 
\acp{NTN} are envisioned to be an essential component of future communication ecosystems~\cite{ericsson2025reportNov}. 
Although \acp{NTN} are typically regarded as auxiliary systems or connectivity providers in rural or remote regions~\cite{giordan2021ntns, geraci2022what},
they can also deliver flexible, on-demand services in urban environments. 
% In particular, aerial \ac{BS} deployment can mitigate \ac{TN} coverage limitations caused by shadowing effects.

This paper presents a comprehensive performance evaluation of \ac{5G} \ac{NTN} realized through aerial \acp{BS} deployed on \ac{LAPS}, 
and compares it with conventional \ac{5G} \ac{TN} in a realistic dense urban environment.
% Unlike battery-limited rotary-wing \acp{UAV}, 
% tethered drones are connected to ground stations via lightweight data and/or power cables,
% enabling potentially unlimited endurance, high-capacity backhaul, and more stable and secure positioning~\cite{fattori2025tethered}. 
% This configuration provides near-\ac{LoS} links to \acp{UE}, mitigates shadowing, and facilitates seamless integration with existing terrestrial \ac{5G} deployments.
% David: I still wonder whether focusing on three UAVs is the right approach, given the energy-consumption concerns, or whether we should instead consider the HAPS concept, which may be more self-sustainable.
% German: while energy-consumption concerns are real, I believe this approach is justified with the following literature review, with several studies where TUAVs are considered as aerial BSs. At the same time, in this paper we consider deployment altitudes significantly below typical HAPS values, but suitable for UAVs.

\subsection{Related Works}

The role of low-altitude platforms --- such as drones and tethered balloons, operating typically below 3000 meters --- in modern communication systems and their potential applications have attracted significant research interest, 
particularly regarding the ability of cellular networks to provide reliable connectivity for \acp{UAV} and the potential of \acp{UAV} to enhance cellular networks~\cite{geraci2022what}.
The later typically motivates the deployment of \ac{UAV}-mounted \acp{BS}. 
However, this approach is constrained by \ac{UAV} energy limitations, 
resulting in restricted payload capacity and flight duration.
To address these limitations, 
\ac{TUAV} \ac{NTN} architectures have been considered in recent studies.
For example, the application of \ac{TUAV} to maximize the coverage of untethered \ac{UAV} relay system was considered in~\cite{safwat2022placement}. 
\Acp{TUAV} have also been employed to enhance coverage in near-shore maritime communications~\cite{ammar2024tethered}.
Zhang et. al.~\cite{zhang2022tethered} proposed a \ac{TUAV}-assisted heterogeneous network,
in which multiple \acp{TUAV} act as aerial relays between \ac{UE} and \ac{TBS}. 
%The authors proposed a cyclic iterative \ac{TUAV} placement, \ac{UE} association and resource allocation framework, aiming to maximize the network sum rate, while accounting for resource constraints, quality of service requirements, and tether entanglement avoidance.
Simulation results demonstrate that the proposed \ac{TUAV} placement, \ac{UE} association and resource allocation framework significantly outperforms a \ac{TBS}-only configuration in terms of aggregate \ac{UE} sum rate.
%Azim {\color{red}please add some of our RL-uav papers if you think they have some to say here}
In~\cite{lim2022optimal}, the authors investigated a \ac{RL}-optimized deployment of \acp{TUAV} for a multi-cell air-to-ground network. 
% The considered system comprises multiple \ac{TUAV}-based \acp{BS} tethered to buildings and operating within their allowable range. 
To maximize network throughput, the authors proposed a multi-agent \ac{RL} framework for controlling \acp{TUAV} positions.
A \ac{RL}-based \ac{UAV} trajectory optimization method was proposed in~\cite{akhtarshenas2026wind}, 
demonstrating the capability of \ac{UAV}-mounted \acp{BS} to operate successfully in GPS-denied scenarios under realistic atmospheric conditions by exploiting \ac{UE} reference signals and angle-of-arrival measurements.
As a trade-off between the sustainability of \acp{TUAV} and the flexibility of untethered \acp{UAV}, 
Cherif~\textit{et al.}~\cite{cherif2022ituavs} proposed the concept of intermittently tethered \acp{UAV}, 
demonstrating that this configuration is particularly effective in scenarios with clustered \ac{UE}.
Later in~\cite{khemirir2023tethered}, the authors proposed a \acp{TUAV} deployment in \ac{UE} hotspots to offload the traffic and assist \acp{TBS}.
%Considering \acp{TUAV} in the large-scale system, the authors evaluated different deployment strategies, including the \ac{TUAV} with the attachment and detachment capability, and analyzed coverage and energy performance for each scenario.
%David: Any positive conclusion regarding the energy performance in this work?
%German: They did not compare energy consumption of TUAV scenarios against terrestrial, if that is what you mean. They only compare energy efficiency of 3 different TUAV scenarios and conclude something well expected: best coverage performance costs more in terms of energy.
Finally, Xu~\textit{et al.}~\cite{xu2025transparent} proposed a rapidly deployable \ac{5G} \ac{TUAV} \ac{BS} with wireless backhaul via a standard \ac{TN} donor \ac{BS} for emergency applications. 
The proposed scheme was validated through prototype field testing.
% \subsection{Motivation and contributions}
% While prior studies primarily consider \ac{TUAV}-based \acp{NTN} as auxiliary or emergency systems and focus on \ac{TUAV} positioning optimization, this work investigates large-scale deployment of \ac{5G} \ac{TUAV}-based \acp{NTN} and compares its performance with conventional \ac{5G} \ac{TN}. We focus on downlink performance and evaluate key system-level metrics, including the average effective \ac{SINR} at the \ac{UE} and the corresponding average \ac{UE} throughput. To the best of the authors’ knowledge, this is the first study to directly evaluate the performance of \ac{5G} \ac{TUAV}-based \ac{NTN} and compare it with conventional \ac{5G} \ac{TN} under a standardized dense urban multi-cell scenario.

\subsection{Motivation and contributions}

Previous studies mainly consider \ac{LAPS}-based \acp{NTN} as auxiliary or emergency systems, focusing on optimizing the positions of a small number of aerial platforms. In contrast, this work examines the large-scale deployment of a \ac{5G} \ac{LAPS}-based \ac{NTN} across a dense urban area with multiple cells and systematically compares its performance with a conventional \ac{5G} \ac{TN} under same deployment conditions.

Rather than optimizing the position of each \ac{LAPS}, here we assess how a common platform altitude across the network affects system-level performance; thereby providing deployment guidelines applicable to realistic large-scale rollouts rather than isolated case studies.

The main contributions of this work are summarized as follows:
\begin{itemize}
    \item We develop a system-level simulation framework for \ac{LAPS}-based \ac{NTN} deployments, extending the standard 3GPP channel model with a geometry-based \ac{LoS} probability model tailored to \ac{LAPS} operating in dense urban environments. 
    Using this extension, we analyze the impact of \ac{LAPS} altitude, ranging from 100\,m to 1000\,m, on network performance, characterizing the trade-off between increased \ac{LoS} probability at higher altitudes and the resulting increase in inter-beam interference.
    \item We evaluate and compare the downlink performance of \ac{LAPS}-based \ac{NTN} and conventional \ac{5G} \ac{TN} deployments under a standardized, 3GPP-compliant multi-cell dense urban scenario, 
    focusing on key system-level metrics, namely the average effective \ac{SINR} at the \ac{UE} and the corresponding average \ac{UE} throughput.
    %\item 
    To the best of the authors' knowledge,
    this is the first study to directly evaluate the performance of \ac{LAPS}-based \ac{NTN} and compare it with conventional \ac{5G} \ac{TN} under a standardized, large-scale dense urban multi-cell scenario.
    \item To mitigate performance degradation caused by inter-beam interference and evaluate the potential of \ac{LAPS}-based \ac{NTN} in dense urban deployments, 
    we proposed a custom beam-steering framework based on a modified antenna pattern modeled according to the TR 38.901 specifications, 
    with an adjustable antenna gain offset.
    %David: I think we should not mention ITU-R M.2135, which is a 4G document, but TR 38.901, which is a 5G document, and has the same mathematical definition for the antenna element.
    %German: agree
    
\end{itemize}

The rest of this paper is organized as follows.
%Section~\ref{sec:RelatedWork} reviews related work on \ac{HAPS}-based \ac{NTN} design and prior comparative analyses.
Section~\ref{sec:SystemModel} describes the system model adopted in this work. 
Section~\ref{sec:Results} presents and discusses the obtained results. 
Finally, Section~\ref{sec:Conclusion} concludes the paper and discusses directions for future research.

%%% Related Work
%\input{Sections/01a_Related_Work}

%%% System Model
\section{System model}\label{sec:SystemModel}

% \begin{figure}[tb]
% \centerline{\includegraphics[width=0.5\textwidth]{Figure/UPA_downtilt.png}}
% \caption{NTN deployment scenario over a hexagonal grid (a), reflector antenna tilt adjustment and HPBW footprint increase for different altitudes (1\,km, 2\,km, 3\,km, and 4\,km for the aperture radius of 10 wavelengths) (b).}
% \label{fig:haps_deployment}
% \end{figure}

In this section,
we describe the system model adopted in this paper.
We consider the downlink performance of terrestrial and non-terrestrial cellular deployments in a \ac{UMa} scenario equipped with \ac{mMIMO} technology,
following the models outlined by 3GPP in~\cite{3GPP38901, 3GPP38811}.

\subsection{Deployment scenarios}

In this work,
we consider three network deployment configurations:

\begin{enumerate}
    \item \textit{TN-UPA}: 
    a conventional terrestrial \ac{5G} deployment with terrestrial \acp{BS} employing a 32-element \ac{UPA} and \ac{DFT}-based beamforming;

    \item \textit{LAPS-UPA}: 
    a non-terrestrial deployment in which the same conventional \ac{5G} \ac{UPA} and beamforming architecture is mounted on \ac{LAPS}, operating at different altitudes;

    \item \textit{LAPS-Custom}: 
    a \ac{LAPS}-based deployment in which the explicit antenna array and beamforming process are replaced by the proposed analytical custom beam-steering model, aimed at mitigating the performance degradation of the \textit{LAPS-UPA} configuration at higher deployment altitudes.
\end{enumerate}
%David: Use these labels to refer to the different cases in the results section.
%German: done

The three configurations share the same carrier frequency, bandwidth, network geometry, and \ac{UE} distribution,
as described in the following.

\begin{table}[!t]
    \centering
    \caption{Simulation parameters}
    \label{tab:param}
    \footnotesize
    \setlength{\tabcolsep}{4pt}
    \renewcommand{\arraystretch}{1.12}

    \begin{tabularx}{\columnwidth}{
        @{} >{\raggedright\arraybackslash}X c r @{}
    }
        \toprule
        \textbf{Parameter} & \textbf{Notation} & \textbf{Value} \\
        \midrule
        Carrier frequency
            & \(f_c\) & 3.5\,GHz \\
        Bandwidth
            & \(B_0\) & 100\,MHz \\
        Inter-site distance
            & \(\mathrm{ISD}\) & 500\,m \\
        Number of site locations
            & --- & 19 \\
        Number of coverage cells
            & --- & 57 \\
        Sector boresight orientations
            & --- & \(30^\circ,\,150^\circ,\,270^\circ\) \\

        \addlinespace[3pt]
        Terrestrial BS altitude
            & \(h_{\mathrm{TN}}\) & 25\,m \\
        Aerial BS altitude
            & \(h_{\mathrm{NTN}}\) & 100--1000\,m \\
        UPA antenna elements
            & --- & \(32\;(4 \times 8)\) \\
        Total BS power
            & \(P^{\mathrm{TX}}\) & 55\,dBm \\

        \addlinespace[3pt]
        Total number of UEs
            & \(N_{\mathrm{UE}}\) & 570 \\
        UE altitude
            & \(h_{\mathrm{UE}}\) & 1.5\,m \\

        \addlinespace[3pt]
        Typical size of dense urban building
            & \(W\) & 40.8\,m \\
        Typical dense urban street width
            & \(S\) & 16.9\,m \\
        \bottomrule
    \end{tabularx}
\end{table}

% \begin{table}[!t]
%     \centering
%     \caption{Simulation parameters}
%     \begin{tabular}{ccc}
%         \textbf{Parameter} & \textbf{Notation} & \textbf{Reference values} \\
%         %4G Carrier frequency [GHz] & \( f_c^{\rm 4G} \) & 2 \\
%         5G Carrier frequency [GHz] & \( f_c^{\rm 5G} \) & 3.5 \\
%         %4G Bandwidth [MHz] & \( B_0^{\rm 4G} \) & 20 \\
%         5G Bandwidth [MHz] & \( B_0^{\rm 5G} \) & 100 \\
%         %Reflector aperture radius [wavelengths] & \(r_{\lambda}\) & 5..50 \\
%         %Reflector aperture radius [m] & \(r_{m}\) & 0.43..4.3 \\
%         TUAV BS altitude [m] & \( h_{\rm NTN} \) & 100..1000 \\
%         TBS altitude [m] & \( h_{\rm TN} \) & 25 \\
%         UE altitude [m] & \( h_{\rm UE} \) & 1.5 \\
%         Total number of UEs & \(N_{\rm UE}\) & 570 \\
%         Inter-site distance [m] & \(ISD\) & 500 \\
%         Typical size of dense urban building [m] & \(W\) & 40.8 \\
%         Typical dense urban street width [m] & \(S\) & 16.9 \\
%         %German: though I agree that it may be good to refer Transmit Power among the parameters, we never mentioned it in the text so placing it in the table is a bit confusing for me.
%         %\red{Transmit Power} & \red{xxx} & \red{xxx} \\
%     \end{tabular}
%     \label{tab:param}
%     % \vspace{-1.5em}
% \end{table}

\subsubsection{Frequency bands}

All considered network configurations operate with bandwidth \(B_0\)
at carrier frequency \(f_c\).
Full frequency reuse is assumed throughout the network.

\subsubsection{Network geometry}

A hexagonal network layout consisting of \(19\) site locations separated by an inter-site distance \(ISD\) is adopted for all network configurations. 
%(see Figure~\ref{fig:network_deployment}).
Each site comprises three sector, with horizontal boresight orientations of \(30^\circ\), \(150^\circ\), and \(270^\circ\), 
resulting in a total of \(57\) cells. 
The \ac{BS} antennas are positioned at different altitudes depending on the considered configuration:
\(h_{\rm TN}\) for the terrestrial \textit{TN-UPA} deployment and
\(h_{\rm NTN}\) for the \ac{LAPS}-based configurations. 
Specifically, the \textit{LAPS-Custom} configuration is considered only for \(h_{\rm NTN}=1000\)\,m.

% \begin{figure}[tb]
%     \centerline{\includegraphics[width=0.35\textwidth]{Figure/network_deployment.png}}
%     \caption{Network deployment over a hexagonal grid}
%     \label{fig:network_deployment}
% \end{figure}

\subsubsection{UE distribution}

We consider \(N_{\rm UE}\) outdoor single-antenna \acp{UE} positioned at a height \(h_{\rm UE}\) and uniformly distributed across all coverage cells. 
This assumption enables a controlled evaluation of network performance under balanced load conditions, 
without introducing additional variability due to non-uniform \ac{UE} spatial distributions.
%David: Are we using indoor UEs? TR38.901 uses indoor UEs. It would have been nice to use them. 
%German: no, right now we only consider outdoor users. Maybe we can consider indoor users as well if we decide to extend this direction in the future.

%In both terrestrial and non-terrestrial deployments, 
%each \ac{UE} is associated with a single serving \ac{BS},
%either ground-mounted in the \ac{TN} scenario or \ac{TUAV}-mounted in the \ac{NTN} scenario.
% David: Cell association is described later. This sentence is confusing. It may mislead the reader into thinking that the two deployments operate simultaneously, i.e., that both networks are deployed at the same time. However, if I understand correctly, you analyze the two scenarios separately, one after the other.
% German: Agree.

\subsection{Antenna and channel models}

In this subsection,
we describe the antenna and channel models adopted for the three considered network configurations.

\subsubsection{LAPS-mounted antenna tilt}

In both the \textit{LAPS-UPA} and \textit{LAPS-Custom} configurations,
the three antennas serving the adjacent sectors of the same site are assumed to be co-located on a single \ac{LAPS}.
Each antenna is mechanically oriented toward the corresponding target service area.
% To this end, the antenna boresight is tilted by an angle \(\alpha\),
% which depends on the \ac{LAPS} altitude and is obtained from the deployment geometry as
% \begin{equation}
%     \alpha = \arctan\!\left(\frac{ISD}{3h_{\rm NTN}}\right).
% \end{equation}

\subsubsection{Antenna configurations}

Following~\cite{3GPP38901},
the \textit{TN-UPA} and \textit{LAPS-UPA} configurations employ a conventional
\ac{5G} \ac{UPA} antenna panel comprising 32 single-polarized elements
arranged in a 4\(\times\)8 grid.
In both configurations, \ac{DFT}-based beamforming is applied, with 8 \ac{SSB} beams and 32 \ac{CSI-RS}.
%David: Are we using 8 SSBs and 32 data beams? Please include this information.
%German: Yes, I included it above.

For these conventional \ac{UPA}-based configurations,
the directional antenna response is obtained from the explicit array model,
accounting for the antenna-element radiation pattern, array geometry,
and beamforming weights.
In contrast, in the \textit{LAPS-Custom} configuration,
the antenna array and beamforming process are not explicitly modeled.
Instead, each beam is represented directly by an analytical radiation pattern
derived from the antenna-element model specified in~\cite{3GPP38901, 3GPP36814},
with parameters controlling its pointing direction and \ac{HPBW}.

\begin{figure}[!t]
    \centering
    \includegraphics[width=1.0\linewidth]{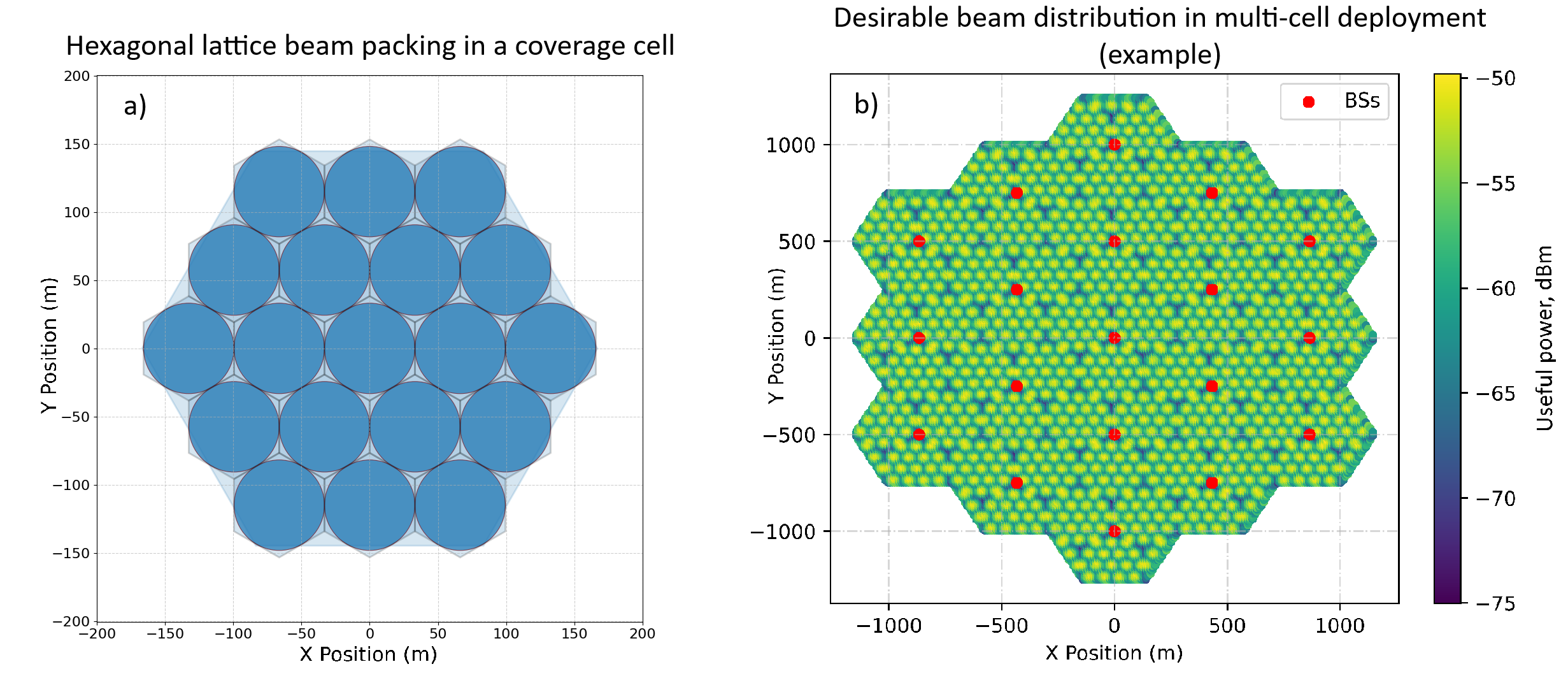}
    \caption{Beam distribution within a single coverage cell (a) and in the multi-cell deployment (b) for the \textit{LAPS-Custom} configuration.}
    \label{fig:hexagonal_packing}
\end{figure}

Specifically, 19 beams are arranged in a hexagonal lattice within each of the 57 coverage cells,
as shown in Figure~\ref{fig:hexagonal_packing}a.
This configuration is designed to reduce inter-beam interference by generating narrow beams directed toward predefined locations within the coverage area while minimizing the overlap between adjacent beam footprints.
Based on the two-dimensional hexagonal deployment geometry,
specific combinations of \ac{LAPS} altitude and \ac{HPBW} are selected to provide efficient coverage with evenly distributed beams while limiting overlap between adjacent beam footprints,
as illustrated in Figure~\ref{fig:hexagonal_packing}b. The \ac{HPBW}, azimuth and elevation angles of each beam are, therefore, explicitly defined by hexagonal lattice geometry and \ac{LAPS} deployment altitude. 

Following the antenna-pattern formulation in~\cite{3GPP38901},
the horizontal (azimuth) and vertical (elevation) radiation patterns,
denoted by $A_{\mathrm{H}}(\varphi)$ and $A_{\mathrm{V}}(\theta)$, respectively,
are defined as follows:
\begin{equation}
    A_{\mathrm{H}}(\varphi)=-\min\left\{12\left(\frac{\varphi}{\varphi_{3\mathrm{dB}}}\right)^{2},A_{\mathrm{m}}\right\},
\end{equation}
and
\begin{equation}
    A_{\mathrm{V}}(\theta)=-\min\left\{12\left(\frac{\theta-\theta_{\mathrm{tilt}}}{\theta_{3\mathrm{dB}}}\right)^{2},A_{\mathrm{m}}\right\},
\end{equation}
where $A_{\mathrm{H}}(\varphi)$ and $A_{\mathrm{V}}(\theta)$ denote the relative horizontal and vertical antenna offset gains, respectively,
$-180^{\circ}\le\varphi\le180^{\circ}$ and $-90^{\circ}\le\theta\le90^{\circ}$ denote the azimuth and elevation angles,
$\varphi_{3\mathrm{dB}}=\theta_{3\mathrm{dB}}$ denotes the \ac{HPBW}, assumed identical in both dimensions,
and $A_{\mathrm{m}}$ denotes the maximum attenuation.
%David: MIMO layer/beam is later defined with l. We should define that in this case L=19 and that the customs angles are function of l.
%German: I extended the paragraph above where we refer Fig 1 and explain beam confguration. The custom angles (both directional and HPBW) are defined by lattice geometry an TUAV altitude. We can provide direct formulas here if you think it is needed, but it may be a bit lengthy.
The combined relative antenna gain is then obtained as
\begin{equation}
    A(\varphi,\theta)
    =
    -\min\bigl\{
    -[A_{\mathrm{H}}(\varphi)+A_{\mathrm{V}}(\theta)],
    A_{\mathrm{m}}
    \bigr\}.
\end{equation}
The corresponding linear directional gain is given by
\begin{equation}
    g(\varphi,\theta)
    =
    10^{\left(G_{\max}+A(\varphi,\theta)\right)/10},
\end{equation}
where \(G_{\max}\) denotes the maximum directional gain in dBi.
In the simulations, \(G_{\max}=42\)~dBi is adopted following the TR 38.901 reference value.

\subsubsection{Channel model}

The channel modeling follows the models outline by 3GPP in~\cite{3GPP38901,3GPP38811},
with adaptations introduced for the \ac{LAPS}-based configurations.

For each \ac{UE} $u$ and cell $c$,
the large-scale propagation gain $\beta_{u,c}$ is defined as
\begin{equation}
    \beta_{u,c} = \rho_{u,c}\tau_{u,c},
\end{equation}
where $\rho_{u,c}$ and $\tau_{u,c}$ denote the path-loss and shadowing gains, respectively.
% \begin{bluenv}
It should be noted that here the shadowing is modeled as a 2D correlated log-normal random variable with zero mean, 
and a standard deviation set according to the link conditions,
i.e., \ac{LoS}/\ac{NLoS}. 
% \end{bluenv}

For the \ac{LAPS}-based configurations,
the path-loss gain $\rho_{u,c}^{\rm NTN}$ is modeled as
%following the ITU-R recommendations in~\cite{ITU-R_P.618-14} as
%David: Before we say we use 3GPP38811 but now ITU-R_P.618-14. Consufing.
%German: indeed. Removed ITU ref as we mentioned 38811 before
\begin{equation}
    \rho_{u,c}^{\rm NTN}
    =
    \frac{1}{
    \rho_{u,c}^{\rm fspl}
    \rho_{u,c}^{\rm cl}
    \rho_{u,c}^{\rm ga}
    \rho_{u,c}^{\rm ra}
    \rho_{u,c}^{\rm ca}
    \rho_{u,c}^{\rm sa}},
\end{equation}
where $\rho_{u,c}^{\rm fspl}$ denotes the free-space propagation loss,
$\rho_{u,c}^{\rm cl}$ denotes the clutter loss,
and $\rho_{u,c}^{\rm ga}$, $\rho_{u,c}^{\rm ra}$,
$\rho_{u,c}^{\rm ca}$, and $\rho_{u,c}^{\rm sa}$
denote the gaseous, rain, cloud, and scintillation attenuation, respectively.

For the \textit{TN-UPA} and \textit{LAPS-UPA} configurations,
the small-scale channel on \ac{PRB} \(k\) is represented by the vector
\(\mathbf{h}_{u,c,k}\), accounting for the explicit antenna array.
For the \textit{LAPS-Custom} configuration,
where the array response is represented analytically through the beam gain,
the corresponding small-scale fading coefficient is denoted by \(h_{u,c,k}\).
In both cases, the small-scale fading follows a Rician model.
\begin{equation}\label{eq:ComplexChannelRician}
    \mathbf{h}_{u,c,k} =
    \sqrt{\frac{K}{1+K}} \, \mathbf{h}^{\rm LOS}_{u,c,k}
    +
    \sqrt{\frac{1}{1+K}} \, \mathbf{h}^{\rm NLOS}_{u,c,k},
\end{equation}
where \(K\) denotes the Rician factor,
\(\mathbf{h}^{\rm NLOS}_{u,c,k}\) represents the \ac{NLoS} channel component modeled as Rayleigh fading,
and \(\mathbf{h}^{\rm LOS}_{u,c,k}\) represents the \ac{LoS} component,
which follows a plane-wave approximation as defined in~\cite{3GPP38901}.
The scalar coefficient \(h_{u,c,k}\) used in the \textit{LAPS-Custom} configuration follows the same Rician formulation.

It is worth noting that the channel model in~\cite{3GPP38811} is defined for altitudes above 8~km.
However, \ac{LAPS}-mounted \acp{BS} operate at substantially lower altitudes,
where the \ac{NLoS} probability may increase due to geometric coverage constraints and stronger interaction with ground-level urban clutter.
To account for this,
we extend the \ac{NTN} channel model by adopting the geometry-based \ac{LoS} probability model proposed by Sabour~\textit{et al.} in~\cite{saboor2024geometry}.
Specifically, the \ac{LoS} probability is calculated as
\begin{equation}
    \mathrm{P}_{\mathrm{LoS}}(\theta, S, W) =
    \frac{S W}{Q}\left( P^{R_1}_{\mathrm{LoS}} + P^{R_2}_{\mathrm{LoS}} \right)
    + \frac{S^{2}}{Q}\, P^{R_3}_{\mathrm{LoS}},
    \label{eq:final_PLOS}
\end{equation}
where $\theta$ denotes the elevation angle,
$P^{R_1}_{\mathrm{LoS}}$ and $P^{R_2}_{\mathrm{LoS}}$ denote the \ac{LoS} probabilities for \acp{UE} located in street regions,
and $P^{R_3}_{\mathrm{LoS}}$ denotes the corresponding probability for \acp{UE} located in crossroad regions.
The parameters $W$ and $S$ denote the typical building size and street width, respectively.
The total street and crossroad area is given by
\begin{equation}
    Q = 2SW + S^2.
\end{equation}
The values of $W$ and $S$ are selected according to~\cite{ITU_R_P1410_6_2023}
for the considered dense urban scenario.
The detailed computation of
$P^{R_1}_{\mathrm{LoS}}$,
$P^{R_2}_{\mathrm{LoS}}$, and
$P^{R_3}_{\mathrm{LoS}}$
is provided in~\cite{saboor2024geometry},
while the values of $W$ and $S$ used in the simulations are reported in Table~\ref{tab:param}.

\subsection{Cell association, SINR and rate computation}

Each \ac{UE} selects the serving cell providing the highest cell-level \ac{RSRP}.
For a given \ac{UE} $u$ and candidate cell $c$,
let $\mathrm{RSRP}_{u,c}$ denote the average received power of the corresponding cell-specific reference signals.
The serving cell is selected as
\begin{equation}
    \hat{c}_u
    =
    \argmax_{c}
    \mathrm{RSRP}_{u,c}.
    \label{eq:CellAssociation}
\end{equation}
Within its serving cell $\hat{c}_u$,
\ac{UE} $u$ is assigned to the transmission beam/layer $\hat{l}_u$ providing the highest beam-level \ac{RSRP}.

Each \ac{UE} is served through a single transmission beam/layer, 
with at most one \ac{UE} scheduled per beam/layer on each \ac{PRB}.
Transmission beams/layers without scheduled \acp{UE} remain inactive.
For the \textit{TN-UPA} and \textit{LAPS-UPA} configurations,
the transmission beams/layers are generated through the explicit \ac{UPA} beamforming process.
For the \textit{LAPS-Custom} configuration,
19 analytically defined beams are available per cell,
i.e., $L=19$.

All configurations assume a fixed total transmit power of $P^{\rm TX}=55~\mathrm{dBm}$ per cell,
uniformly allocated across the available \acp{PRB} and transmission beams/layers.
The transmit power allocated to beam/layer $l$ of cell $c$ on \ac{PRB} $k$ is denoted by $p^l_{c,k}$,
while $\sigma_k^2$ denotes the corresponding noise power.

The downlink \ac{SINR} is computed differently for the
\ac{UPA}-based and \textit{LAPS-Custom} configurations,
owing to their different antenna models.

For the \textit{TN-UPA} and \textit{LAPS-UPA} configurations,
the directional gain of an individual antenna element is denoted by $g^{\rm el}_{u,c}$,
while the array and beamforming effects are explicitly captured through the channel and precoding vectors.
The corresponding downlink \ac{SINR} is given by \eqref{eq:SINR_UPA},
where $\mathcal{U}_c$ denotes the set of \acp{UE} served by cell $c$ and $\mathbf{w}^{\hat{l}_u}_{u,c,k}$ denotes the precoding vector associated with \ac{UE} $u$.

For the \textit{LAPS-Custom} configuration,
the explicit antenna array and precoding process are replaced by an analytical beam model.
The directional gain $g^{\mathrm{beam}}_{u,c,l}$ represents the gain experienced by \ac{UE} $u$ from beam $l$ of cell $c$.
The corresponding downlink \ac{SINR} is given by \eqref{eq:SINR_Custom},
where the first and second interference terms represent intra-cell and inter-cell interference, respectively.

For all considered network configurations,
the effective \ac{SINR} of \ac{UE} \(u\) on its serving transmission beam/layer \(\hat{l}_u\),
denoted by \(\tilde{\gamma}_{u,\hat{l}_u}\),
is obtained from the corresponding set of per-\ac{PRB} \acp{SINR} \(\gamma_{u,\hat{l}_u,k}\) using a mutual-information-based effective-\ac{SINR} mapping.
This effective \ac{SINR} provides an overall representation of the link quality experienced by the \ac{UE} over its allocated frequency resources.

Assuming round-robin scheduling and full-buffer traffic,
the achievable downlink throughput of \ac{UE} \(u\) is computed as
\begin{equation}
    R_u =
    \frac{
        N^{\rm PRB}_{\hat{c}_u}
        B^{\rm PRB}_{\hat{c}_u}
    }{
        N^{\rm UE}_{\hat{c}_u,\hat{l}_u}
    }
    \log_2\!\left(
        1+\tilde{\gamma}_{u,\hat{l}_u}
    \right),
    \label{eq:AchievableDataRate}
\end{equation}
where \(N^{\rm PRB}_{\hat{c}_u}\) is the total number of \acp{PRB} available at the serving cell \(\hat{c}_u\),
\(B^{\rm PRB}_{\hat{c}_u}\) is the bandwidth of each \ac{PRB},
and \(N^{\rm UE}_{\hat{c}_u,\hat{l}_u}\) is the number of \acp{UE} sharing transmission beam/layer \(\hat{l}_u\) of the serving cell under round-robin scheduling.

\begin{strip}
\small
    \begin{equation}
        \gamma_{u,\hat{l}_u,k}^{\rm UPA}
        =
        \frac{
            \beta_{u,\hat{c}_u}
            g^{\rm el}_{u,\hat{c}_u}
            \left|
            \mathbf{h}_{u,\hat{c}_u,k}
            \mathbf{w}^{\hat{l}_u}_{u,\hat{c}_u,k}
            \right|^2
            p^{\hat{l}_u}_{\hat{c}_u,k}
        }{
            \sum\limits_{ u' \in \mathcal{U}_{\hat{c}} \setminus u}  
            \beta_{u,\hat{c}_u}
            g^{\rm el}_{u,\hat{c}_u}
            \left|
            \mathbf{h}_{u,\hat{c}_u,k}
            \mathbf{w}^{\hat{l}_{u'}}_{u',\hat{c}_u,k}
            \right|^2
            p^{\hat{l}_{u'}}_{\hat{c}_u,k}
            +
            \sum\limits_{c \neq \hat{c}_u}
            \sum\limits_{u' \in \mathcal{U}_c}
            \beta_{u,c}
            g^{\rm el}_{u,c}
            \left|
            \mathbf{h}_{u,c,k}
            \mathbf{w}^{\hat{l}_{u'}}_{u',c,k}
            \right|^2
            p^{\hat{l}_{u'}}_{c,k}
            +
            \sigma_k^2
        }
        \label{eq:SINR_UPA}
    \end{equation}

    \begin{equation}
    \gamma^{\mathrm{Custom}}_{u,\hat{l}_u,k}
    =
    \frac{
    \beta_{u,\hat{c}_u}
    g^{\mathrm{beam}}_{u,\hat{c}_u,\hat{l}_u}
    |h_{u,\hat{c}_u,k}|^2
    p^{\hat{l}_u}_{\hat{c}_u,k}
    }{
    \displaystyle
    \sum_{u'\in\mathcal{U}_{\hat{c}_u}\setminus u}
    \beta_{u,\hat{c}_u}
    g^{\mathrm{beam}}_{u,\hat{c}_u,\hat{l}_{u'}}
    |h_{u,\hat{c}_u,k}|^2
    p^{\hat{l}_{u'}}_{\hat{c}_u,k}
    +
    \sum_{c\neq\hat{c}_u}
    \sum_{u'\in\mathcal{U}_c}
    \beta_{u,c}
    g^{\mathrm{beam}}_{u,c,\hat{l}_{u'}}
    |h_{u,c,k}|^2
    p^{\hat{l}_{u'}}_{c,k}
    +\sigma_k^2
    }.
        \label{eq:SINR_Custom}
    \end{equation}    
\end{strip}

%%% Results
\section{Simulation Results}\label{sec:Results}

In this section, 
we present the main results of the performance evaluation of the three network configurations introduced in Section~\ref{sec:SystemModel}, 
\textit{TN-UPA}, \textit{LAPS-UPA}, and \textit{LAPS-Custom}.

\begin{figure}[!t]
    \centering
    \includegraphics[width=1.0\linewidth]{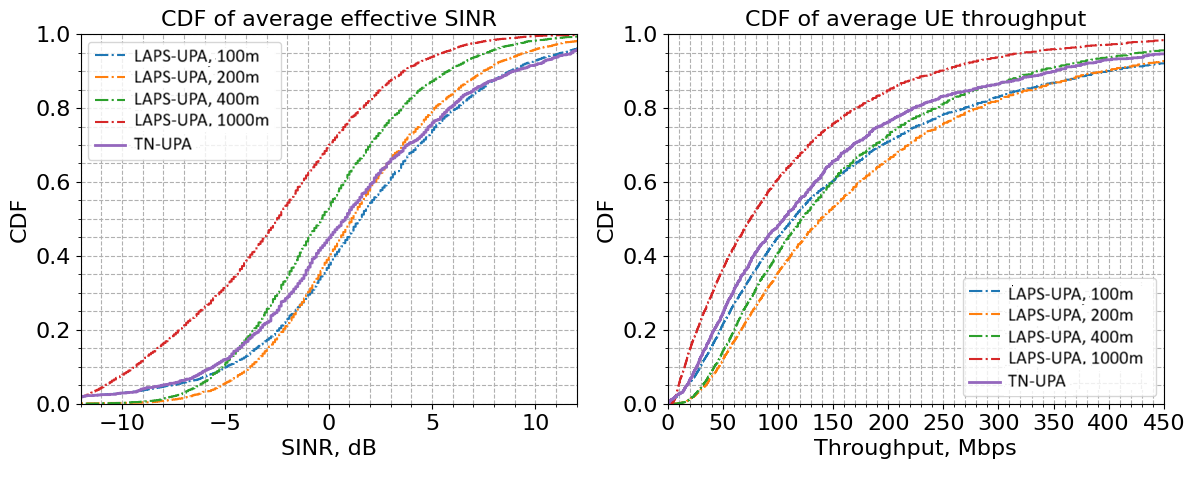}
    \caption{Average effective SINR and UE throughput for the \textit{TN-UPA} and \textit{LAPS-UPA} configurations.}
    \label{fig:performance}
% \vspace{-1em}
\end{figure}

The experiments are conducted using \textit{Giulia}, 
a system-level simulator calibrated according to 3GPP deployment and channel models,
to provide a realistic network-level performance evaluation\footnote{\texttt{Giulia} is publicly available at \url{https://github.com/giulia-open-lab/OpenGiuliaSLS.git}.}.
The main simulation parameters are summarized in Table~\ref{tab:param}.

We first evaluate the \textit{LAPS-UPA} configuration and investigate the impact of the \ac{LAPS} altitude,
ranging from 100\,m to 1000\,m.
The resulting distributions of the effective \ac{SINR} and \ac{UE} throughput are shown in Figure~\ref{fig:performance},
together with the corresponding results for the benchmark \textit{TN-UPA} configuration.
The results show a clear dependence of network performance on the \ac{LAPS} altitude.
In particular, at \(h_{\rm NTN}=200\)~m,
the median \ac{UE} throughput of the \textit{LAPS-UPA} configuration reaches approximately 140\,Mbps,
compared with approximately 100\,Mbps for \textit{TN-UPA}.
This value also exceeds the performance obtained at both lower and higher \ac{LAPS} altitudes.
Similarly, the 5th-percentile \ac{UE} throughput reaches approximately 30\,Mbps at \(h_{\rm NTN}=200\)~m,
compared with approximately 20\,Mbps for the \textit{TN-UPA} configuration.

\begin{figure}[!t]
    \centering
    \includegraphics[width=1.0\linewidth]{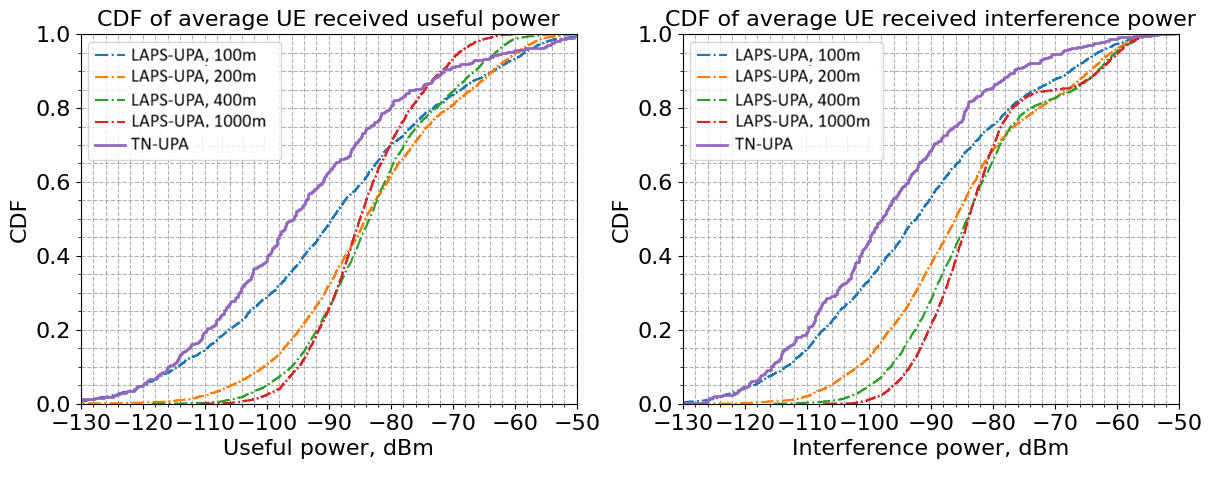}
    \caption{Average useful and interference powers received by the UEs for the \textit{TN-UPA} and \textit{LAPS-UPA} configurations.}
    \label{fig:power}
% \vspace{-1.5em}
\end{figure}

To better understand why the \textit{LAPS-UPA} configuration achieves its best performance around \(h_{\rm NTN}=200\)~m,
we analyze the average useful and interference powers received by the \acp{UE},
as shown in Figure~\ref{fig:power}.
Both quantities are generally higher for \textit{LAPS-UPA} than for \textit{TN-UPA},
as the elevated deployment increases the \ac{LoS} probability and modifies the interference geometry.

The most pronounced improvement occurs between 100\,m and 200\,m.
Considering the target coverage areas % shown in Figure~\ref{fig:network_deployment},
with an approximate cell diameter of 330\,m and an \(ISD\) of 500\,m,
the minimum elevation angle of a served \ac{UE} increases from approximately \(17^\circ\) at 100\,m to \(31^\circ\) at 200\,m.
According to the adopted geometry-based model~\cite{saboor2024geometry},
this increase approximately doubles the \ac{LoS} probability from 0.2 to 0.4.
%as illustrated in Figure~\ref{fig:plos}.
The resulting improvement in propagation conditions explains the significant increase in useful received power and,
consequently,
the higher effective \ac{SINR} and throughput observed at 200\,m.

% \begin{figure}[!b]
%     \centering
%     \includegraphics[width=0.6\linewidth]{Figure/PLoS.png}
%     \caption{Analytical estimation of LoS probability for a dense urban scenario}
%     \label{fig:plos}
% \vspace{-1.5em}
% \end{figure}

Further increasing the altitude,
however,
does not provide the same benefit.
Although the \ac{LoS} probability continues to increase,
the longer propagation distance also increases the path loss,
while the ground footprints of the \ac{DFT} beams become larger.
The latter, in turn, increases the overlap between neighboring beam footprints and,
therefore,
the interference experienced by the \acp{UE}.
As illustrated in Figure~\ref{fig:power},
at high altitudes the increase in interference progressively offsets the propagation benefits obtained from the higher \ac{LoS} probability.
As a result, the \textit{LAPS-UPA} configuration achieves its best overall performance around 200\,m,
while performance degrades again as the platform altitude approaches 1000\,m.

Motivated by this observation,
we next evaluate the \textit{LAPS-Custom} configuration,
which employs narrower analytically defined beams to reduce inter-beam interference.
To evaluate the capability of the \textit{LAPS-Custom} configuration to mitigate the performance degradation observed with the \textit{LAPS-UPA} configuration, the former was simulated at an altitude of 1000\,m (where the latter exhibited a significant performance drop), using \ac{HPBW} angles  \(\varphi_{3\mathrm{dB}}=\theta_{3\mathrm{dB}}=2^\circ\).
At this altitude and beamwidth,
the 19 beams provide the hexagonal coverage arrangement illustrated in Figure~\ref{fig:hexagonal_packing}a.
We consider the reference maximum attenuation \(A_{\mathrm{m}}=20\)~dB,
as well as values of 40, 60, and 80~dB.
Increasing \(A_{\mathrm{m}}\) strengthens the suppression of off-beam radiation,
thereby reducing interference between simultaneously active beams.

\begin{figure}[!t]
    \centering
    \includegraphics[width=1.0\linewidth]{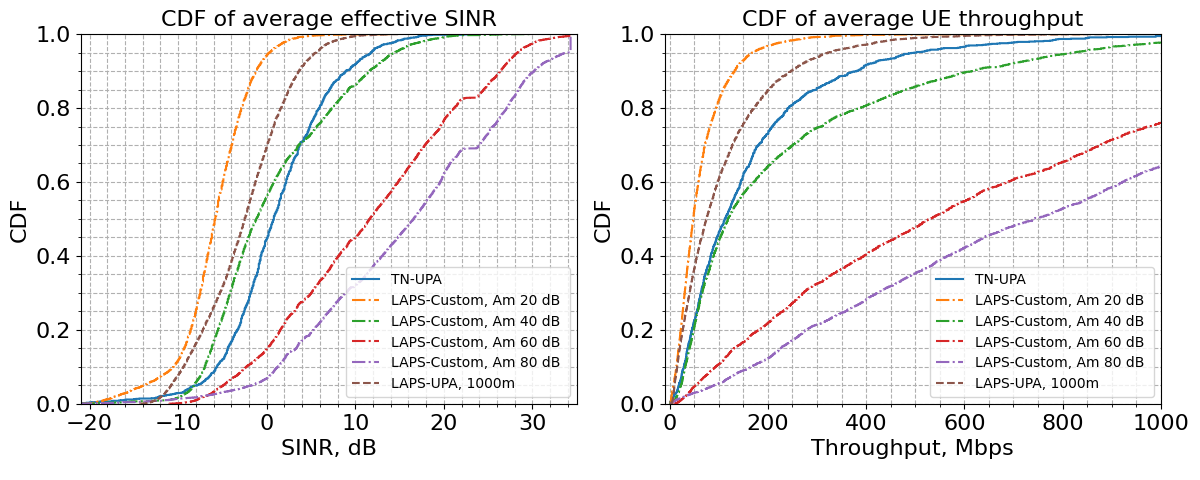}
    \caption{Average effective SINR and UE throughput for the \textit{TN-UPA}, \textit{LAPS-UPA} at 1000\,m altitude, and \textit{LAPS-Custom} configurations.}
    \label{fig:custom_performance}
% \vspace{-1.5em}
\end{figure}

\begin{figure}[!t]
    \centering
    \includegraphics[width=1.0\linewidth]{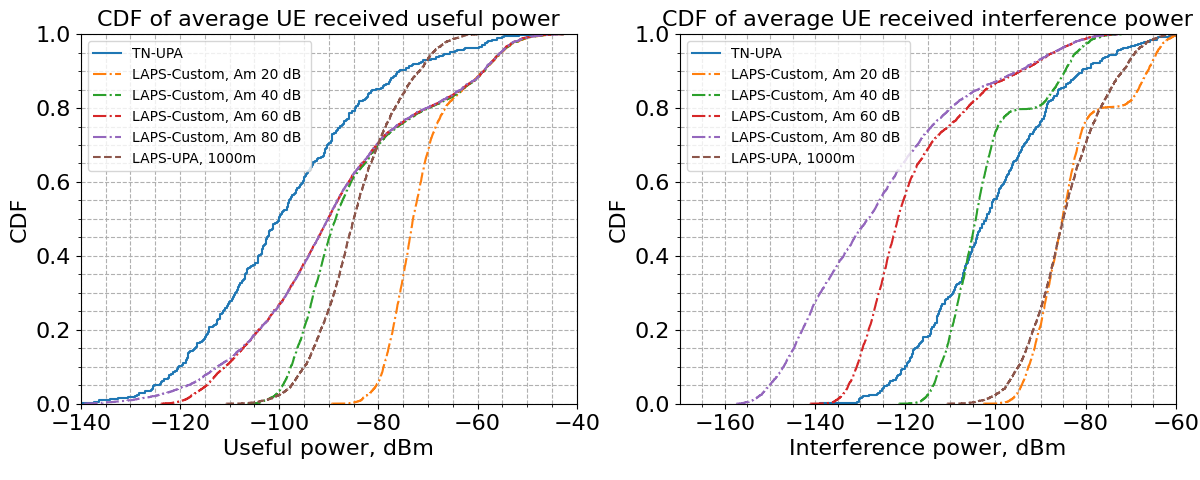}
    \caption{Average useful and interference powers for the \textit{TN-UPA}, \textit{LAPS-UPA} at 1000\,m altitude, and \textit{LAPS-Custom} configurations.}
    \label{fig:custom_power}
\vspace{-1.5em}
\end{figure}

Figure~\ref{fig:custom_performance} compares the effective \ac{SINR} and \ac{UE} throughput,
while Figure~\ref{fig:custom_power} reports the corresponding useful and interference powers.
The results show that stronger off-beam suppression substantially improves the performance of the \textit{LAPS-Custom} configuration.
The median effective \ac{SINR} increases from approximately \(-3\)~dB for \textit{LAPS-UPA} at 1000\,m and approximately 1~dB for \textit{TN-UPA},
to approximately 12~dB and 16~dB for \textit{LAPS-Custom} with \(A_{\mathrm{m}}=60\) and 80~dB correspondingly.
Similarly,
the median \ac{UE} throughput increases from approximately 75\,Mbps for \textit{LAPS-UPA} and 120\,Mbps for \textit{TN-UPA},
to approximately 525\,Mbps and 750\,Mbps for the corresponding \textit{LAPS-Custom} configurations.
At the same time, for the \textit{LAPS-Custom} with \(A_{\mathrm{m}}=20\) and 40~dB the median effective \ac{SINR} remain below the \textit{TN-UPA} scenario with approximate values of \(-6\) and \(-1\)~dB correspondingly.
Similarly,
the median \ac{UE} throughput is limited by 50 and approximately 120\,Mbps for the corresponding \textit{LAPS-Custom} configurations.
%David: I am missing the KPI values that we obtain with a suprresion of 40, 60, and 80~dB.
%German: added results for 20 and 40 dB also

Considering~\eqref{eq:AchievableDataRate} and the fact that all scenarios operate with the same bandwidth and total number of available \acp{PRB}, the median achievable throughput gain can be attributed to two factors: increased \ac{SINR} and more efficient resource allocation. In the \textit{LAPS-UPA} and \textit{TN-UPA} scenarios, some DFT beams may point outside the target area, contributing less effectively to coverage. In contrast, the \textit{LAPS-Custom} scenario uses fewer beams but distributes them more uniformly across the coverage area, which is advantageous for a balanced, non-clustered user distribution.
%The numerical impact of this effect can be estimated from the throughput comparison of \textit{LAPS-Custom} scenarios with \(A_{\mathrm{m}}=60\) and 80~dB against the \textit{TN-UPA} scenario. The \ac{SINR} gain alone between median \textit{TN-UPA} \ac{SINR} (1~dB) and \textit{LAPS-Custom} \ac{SINR} under \(A_{\mathrm{m}}=60\) (12~dB) will give about 3.7x increase in logarithmic multiplier in the~\eqref{eq:AchievableDataRate}.

The improvement in median \ac{SINR} and \ac{UE} throughput in \textit{LAPS-Custom} configurations with higher maximum attenuation \(A_{\mathrm{m}}\) is primarily associated with the reduction in interference.
For \(A_{\mathrm{m}}=60\) and 80~dB,
the mean interference power decreases to approximately \(-122\) to \(-128\)~dBm,
compared with approximately \(-85\)~dBm for \textit{LAPS-UPA} at 1000\,m and \(-102\)~dBm for \textit{TN-UPA}.
However, 
stronger off-beam suppression also reduces the useful power received by some \acp{UE}, particularly near the boundaries between adjacent beam footprints. 
% Meanwhile, 
% the increased propagation distance relative to the terrestrial scenario and the resulting propagation losses further reduce the received useful power, requiring a higher total \ac{BS} power. 
Consequently, 
the 5th-percentile effective \ac{SINR} and throughput gain of the \textit{LAPS-Custom} configurations remain limited, highlighting a trade-off between interference suppression and coverage continuity.

%%% Conclusion
\section{Conclusion and Future Work}\label{sec:Conclusion}

This paper evaluated \ac{LAPS}-based \ac{5G} \ac{NTN} deployments against conventional terrestrial \ac{5G} in a dense urban scenario,
considering platform altitudes ranging from 100\,m to 1000\,m.
The results show that aerial deployments can outperform conventional terrestrial networks when the \ac{LAPS} altitude and antenna configuration are appropriately selected.
For the \textit{LAPS-UPA} configuration,
an intermediate altitude provides the best trade-off between \ac{LoS} conditions and interference.
The \textit{LAPS-Custom} results further demonstrate that interference suppression is a key enabler for aerial deployments, as appropriate beam shaping and off-beam radiation suppression can substantially improve performance even at higher altitudes, although this may require a capability to generate significantly more focused beams and, therefore, deploy bigger antenna arrays.
These results indicate that interference-aware aerial network design represents a promising direction for future cellular networks.

As future work,
we plan to investigate the joint optimization of \ac{LAPS} placement and beam configuration,
with particular emphasis on interference-aware beam management using deep reinforcement learning.
We will also investigate non-uniform \ac{UE} distributions, \ac{UE} mobility, and multi-\ac{UAV} coordination,
with the broader objective of enabling efficient integration of aerial base stations into future cellular networks.

\bibliographystyle{ieeetr}
\bibliography{journalAbbreviations, refs}

\end{document}